\documentclass[aps,twocolumn,amsmath,amssymb,superscriptaddress,prl,showpacs]{revtex4}
\usepackage{amsfonts, hyperref}
\usepackage{graphicx}
\pdfoutput=1

\begin{document}

\title{Large thermoelectric figure of merit for 3D topological
  Anderson insulators via line dislocation engineering}

\author{O.~A.~Tretiakov}
\affiliation{
            Department of Physics,
            MS 4242,
	    Texas A\&M University,
            College Station, TX 77843-4242, USA
}
\author{Ar. Abanov}
\affiliation{
            Department of Physics,
            MS 4242,
	    Texas A\&M University,
            College Station, TX 77843-4242, USA
}

\author{Shuichi Murakami}
\affiliation{Department of Physics, Tokyo Institute of Technology,
  Ookayama, Meguro-ku, Tokyo 152-8551, Japan}
\affiliation{PRESTO, Japan Science and Technology Agency (JST),
  Kawaguchi, Saitama 332-0012, Japan}

\author{Jairo Sinova}
\affiliation{
            Department of Physics,
            MS 4242,
	    Texas A\&M University,
            College Station, TX 77843-4242, USA
}

\date{July 17, 2010}

\begin{abstract}
We study the thermoelectric properties of three-dimensional
topological Anderson insulators with line dislocations. We show that
at high densities of dislocations the thermoelectric figure of merit
$ZT$ can be dominated by one-dimensional topologically-protected
conducting states channeled through the lattice screw dislocations in
the topological insulator materials with a non-zero
time-reversal-invariant momentum such as $\rm{Bi}_{1-x}\rm{Sb}_x$.
When the chemical potential does not exceed much the mobility edge the
$ZT$ at room temperatures can reach large values, much higher than
unity for reasonable parameters, hence making this system a strong
candidate for applications in heat management of nano-devices.
\end{abstract}


\maketitle

\textit{Introduction.}  The recent crisis of heat management in
nano-devices, which has lead to a lack of progression in clock speeds
of charge-based logic devices, has intensified the interest in
efficient thermoelectric materials.  In the last decade there has been
a lot of research both theoretical \cite{DiVentra09, Murakami2010,
  Markussen09, mukerjee07, Ghaemi10} and experimental
\cite{venkatasubramanian01, Ho-KiLyeo04, zhang:062107} to create
efficient thermoelectric nano-devices.  The efficiency of such
materials, which must be both p-type and n-type, is determined by a
balance to convert charge flow into efficient heat transport as well
as maintaining a temperature gradient between the device and the heat
sink.  Among the most well known thermoelectric materials in present
day commercial applications one finds Bi$_2$Te$_3$, PbTe, and
PbSb. These type of insulators or semiconductors have been identified
recently as topological insulators (TI) \cite{TI_physics_today} which
exhibit protected delocalized surface states.

In the two-dimensional version of the TIs, the quantum spin Hall
systems \cite{Bernevig06, Molenkamp07}, these protected edge states
contribute to the thermoelectric efficiency but do not enhance it
dramatically beyond present day materials whose efficiency parameter,
$ZT$ (see below), is of 1 or slightly less \cite{Murakami2010}.  On
the other hand, for three-dimensional (3D) topological insulators with
a non-zero time-reversal-invariant momentum \cite{Fu07} (TRIM) it has
been shown theoretically that one-dimensional (1D) topologically
protected modes can exist in the bulk propagating through certain line
dislocations \cite{ran_one-dimensional_2009}.  These type of 1D
quantum modes have been recently attributed to the results of recent
experiments on $\rm{Bi}_{2}\rm{Se}_3$ \cite{Checkelsky09, Hor09}.

\begin{figure}
\includegraphics[width=1.0 \columnwidth]{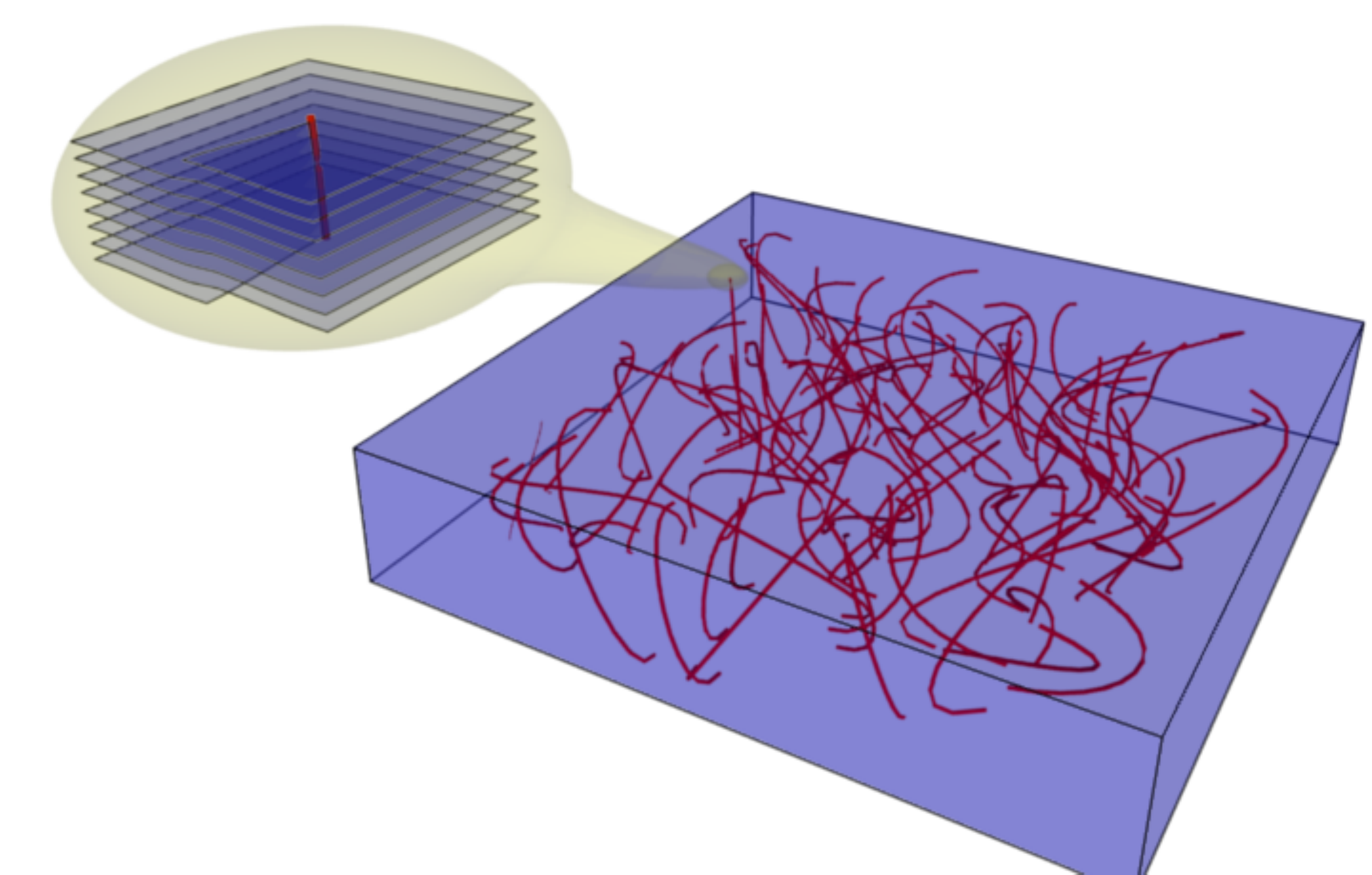} 
\caption{(Color online) A sketch of 3D topological insulator with
  lattice dislocations propagating through it. These dislocations are
  topologically-protected 1D modes which are perfectly conducting. The
  inset shows the details of the screw-type dislocation considered.}
\label{fig:dislocations} 
\end{figure}

Here we explore the idea of using these 1D topologically protected
modes to significantly increase the thermoelectric efficiency of
materials such that $\rm{Bi}_{1-x}\rm{Sb}_x$, see
Fig.~\ref{fig:dislocations}.  The basic premise of the proposal is to
introduce, through growth engineering, a finite density of screw
dislocations. This would induce disorder in the bulk leading to a
reduction of the thermal conductivity, Anderson localization of bulk
states, and an increase of the conductivity and thermopower
contributions from these 1D states. This combination of factors, as
shown below, leads to a dramatic enhancement of the figure of merit
efficiency for thermoelectrics, $ZT$, beyond its present value for
bulk materials.  For reasonable parameters we estimate $ZT$ to reach
$\sim 6$ at room temperature.

\textit{Figure of merit}. The performance of thermoelectric devices is
determined by the dimensionless figure of merit $ZT$ defined as
\begin{equation}
ZT = \frac{\sigma S^2 T}{\kappa},
\label{ZT1}
\end{equation}
where $\sigma$, $S$, $\kappa$, and $T$ are the electrical
conductivity, Seebeck coefficient (or thermopower), thermal
conductivity, and absolute temperature, respectively. The thermal
conductivity generally has two contributions: electronic and phononic,
$\kappa =\kappa_{e}+\kappa_{ph}$. A large effort within thermoelectric
studies is devoted to finding materials with high $ZT$ near room
temperature. The challenge is that effects that increase $\sigma$ or
$S$ usually accompanies  an increase on $\kappa$ and vice versa. In
particular, an important challenge is to reduce the phononic thermal
contribution to $\kappa$ which can limit severely $ZT$. Doing so
without reducing at the same time the $\sigma S^2 T$ factor is a key
challenge.

Our proposal is to exploit the 1D topological protected channels
formed at the dislocations of a TI with non-zero TRIM
\cite{ran_one-dimensional_2009} to circumvent this challenge.  As more
dislocations are introduced in the system the mean free path of the
phonons is reduced while at the same time the conductivity
contribution from the 1D channels is increased as mentioned above.  By
requiring that the bulk contribution is also reduced by Anderson
localization of the bulk states, through the dislocations themselves
or other co-doping, the $ZT$ factor will be dominated exclusively by
the 1D channels and a large $ZT$ value can be obtained
\cite{conference}. Below we present our theoretical estimate for this
high $ZT$ using reasonable estimates of the different parameters
without seeking the best possible scenario but estimating instead a
reasonable expectation of the proposed system.

Within linear response theory \cite{AshcroftMermin} the electric
($j^e$) and thermal ($j^q$) currents are given by linear combinations
of the chemical potential and temperature gradients: $j^e/e= L_0\nabla
\mu + L_1 (\nabla T)/T$ and $j^q= -L_1\nabla \mu - L_2 (\nabla T)/T$,
where $e$ is the electron charge.  From these equations, using Onsager
relations, one can find that the electrical conductivity $\sigma= e^2
L_0$, thermopower $S=-L_1/(eTL_0)$, and electronic thermal
conductivity $\kappa_{e}=(L_0 L_2 -L_1^2)/(TL_0)$. Then according to
Eq.~\eqref{ZT1} in terms of transport coefficients $L_\alpha$ the
figure of merit takes the form \cite{marder}:
\begin{widetext}
\begin{equation}
ZT= \frac{(L_1^b+ sn L_1^{1D})^2}{(L_0^b+ sn L_0^{1D})(L_2^b+ sn L_2^{1D})
-(L_1^b+ sn L_1^{1D})^2 + 
\kappa_{ph}(L_0^b+ sn L_0^{1D})T},
\label{ZT2}
\end{equation}
\end{widetext}
where $n$ is the density of topologically-protected lattice
dislocations \cite{ran_one-dimensional_2009}, $s$ is the
cross-sectional area of the device transverse to the transport
direction, and $\kappa_{ph} =\frac{c_{ph}}{3}v_{ph}l_{ph}$ is the
phonon contribution to the thermal conductivity. This expression for
$\kappa_{ph}$ is applicable at room temperatures.  In Eq.~\eqref{ZT2}
it is assumed that the transport coefficients have bulk and 1D channel
contributions $L_\alpha = L^b_\alpha + snL_\alpha^{1D}$, where $sn$
gives the number of the topologically protected lattice dislocations
conducting perpendicular to area $s$.

Next we take $l_{ph}$ in the limit of high density of randomly located
dislocations, such that for high enough dislocation density $n$,
$l_{ph}$ is diminished by phonon scattering from these dislocations.
The phonon specific heat $c_{ph}$ at room temperature can be estimated
as $3n_i k_B$, where the number of ions per unit volume is $n_i \sim
4\cdot 10^{29}\, \rm{m}^{-3}$. In (Bi$_{1-x}$Sb$_{x}$)$_{2}$Te$_{3}$
and Bi$_{2}$(Te$_{1-y}$Se$_{y}$)$_{3}$ compounds the phonon velocity
$v_{ph}=1500$ m$/$s and Debye temperature $\theta_D =142$ K
\cite{Yokota73}.  We also take into account in our calculations the
fact that the bulk becomes an amorphous media at very high $n$ due to
disorder which leads to the saturation of $\kappa_{ph}$.  Therefore,
$\kappa_{ph}$ ranges from $1 \rm{Wm}^{-1}\rm{K}^{-1}$ for pure bulk
with no dislocations \cite{Pattamatta09} to $\kappa_{ph} \approx 0.01
\rm{Wm}^{-1}\rm{K}^{-1}$ at average distances between dislocations
$d\sim 3$ nm.  It is clear from Eq.~\eqref{ZT2} that at large
densities of lattice dislocations, $n$, the contribution to $ZT$
mostly comes from 1D channels and in the limit equals to $ZT$ of one
perfectly conducting 1D wire:
\begin{equation}
\lim_{n\rightarrow \infty} ZT= \frac{(L_1^{1D})^2}{L_0^{1D}L_2^{1D} -(L_1^{1D})^2},
\label{ZT_1D}
\end{equation}

\begin{figure}
\includegraphics[width=0.99\columnwidth]{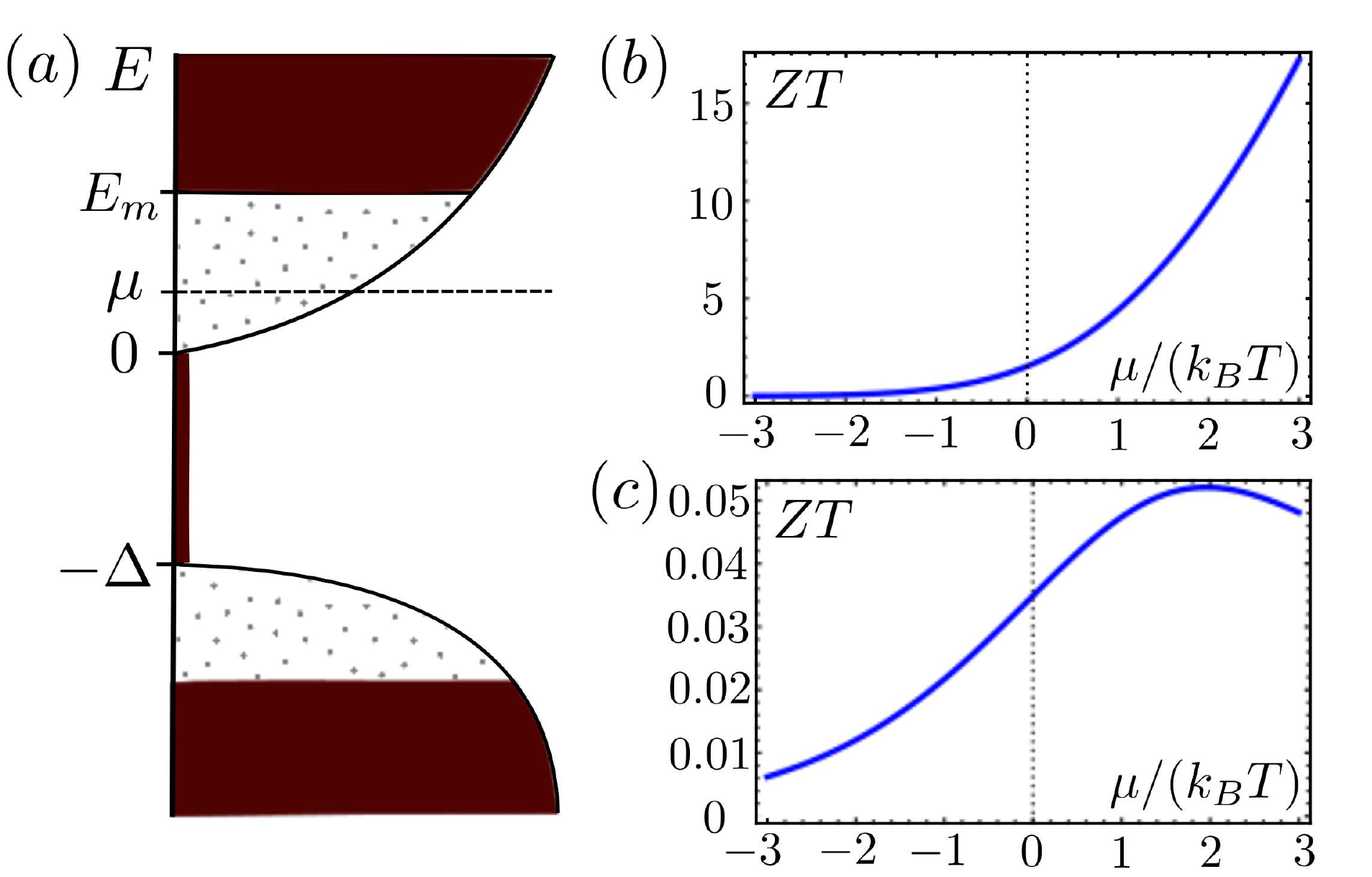}
\caption{(Color online) (a) Schematic band structure of the bulk and
  1D states. (b) Figure of merit $ZT$ for the contribution from
  topologically-protected 1D channels as a function of chemical
  potential $\mu$ (in units $k_B T$) at temperature $T = 300$ K. (c)
  Pure bulk contribution to $ZT$ as a function of $\mu$.}
\label{fig:band_structure}
\end{figure}

To make estimates of the relative contributions of the 1D channels and
bulk we model the topological Anderson insulator system as a
semiconductor with one valence and one conduction band and one
1D-state corresponding to a perfectly conducting lattice dislocation
(in general, one 1D-state per each dislocation), see
Fig.~\ref{fig:band_structure} (a).  We take the gap between conduction
and valence bands in the bulk to be $\Delta$ ($-\Delta<E<0$) and the
transmission coefficient in the 1D channel is assumed to be unity
${\cal T}(E) =1$ for simplicity. Generally $\Delta= 0.15$ eV for
$\rm{Bi}_{2}\rm{Te}_3$ and $\Delta= 0.3$ eV for $\rm{Bi}_{2}\rm{Se}_3$
we use the former for our estimates.  The bulk states near the edges
of the band are localized with a mobility edge at $E_m$; this Anderson
bulk localization is due to the high density of dislocations or can be
induced by doping with non-magnetic impurities.

Employing the Landauer-Buttiker formalism \cite{Sivan86, Butcher90} we
can write the expressions for the transport coefficients in the 1D
channel:
\begin{equation}
L_\alpha^{1D} = - \frac{l}{sh}\int\!\! {\cal T}(E) f' (E)(E-\mu)^{\alpha}dE ,
\end{equation}
where $l$ is the length of the sample in the growth direction (length
of 1D channel) which has the upper limit of the inelastic coherence
length ($l_{in}\sim 1 \mu{\rm m}$), $h$ is a Planck constant, and $f'
(E)=\partial f/\partial E$ with $f=1/(e^{(E-\mu)/(k_BT)}+1)$ being the
Fermi function. Here the integration over energies extends from
$-\Delta$ to $0$ while the chemical potential, which can be changed by
an external gate, we restrict to be in the gap, $-\Delta<\mu<0$, or
within the conduction band, $\mu >0$, but not much above the mobility
edge. The latter case is of the most interest, since in this case $ZT$
is the largest, see Fig.~\ref{fig:band_structure} (b).

\begin{figure}
\includegraphics[width=0.99\columnwidth]{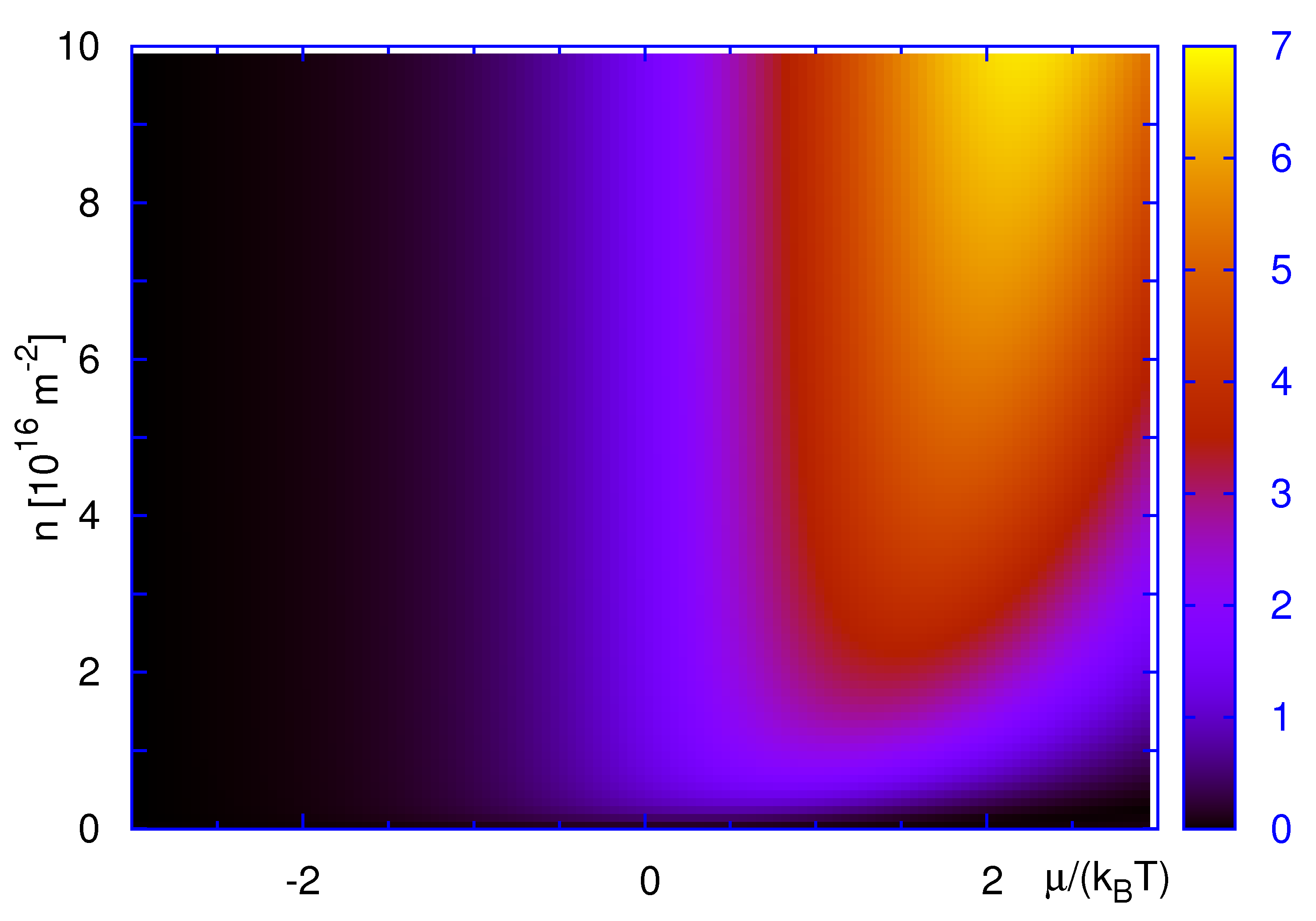}
\caption{(Color online) The contour plot of figure of merit $ZT$ as a
  function of chemical potential $\mu$ (in units $k_B T$) and the
  density of lattice dislocations $n$ at temperature $T = 300$ K for
  sample's length $l=1 \mu{\rm m}$. The maximum value is higher than
  6.}
\label{fig:ZT2}
\end{figure}

To estimate the bulk contribution to the transport coefficients
$L_\alpha$ we assume the bands to be parabolic and use Boltzmann
equation in the relaxation time approximation,
\begin{equation}
L_\alpha^b = -\tau \int_{E_m}^{\infty}  D(E)  f'(E) v^2 (E-\mu)^{\alpha}dE.
\label{Lbulk}
\end{equation}
Here $v$ is electron velocity, $D(E)$ is the density of states, and we
approximate relaxation time $\tau$ to be independent of energy.  The
extended and localized states are separated by the mobility edge $E_m$
which is measured from the bottom of the conduction band, see
Fig.~\ref{fig:band_structure} (a), and for further estimates we assume
$E_m =0.05$ eV. Then Eq.~\eqref{Lbulk} gives
\begin{equation}
L_\alpha^b =\frac{2\sqrt{2m^{*}}}{\pi^2\hbar^3} \tau c T^{\alpha +3/2} 
\int^{\infty}_{\frac{E_m -\mu}{T}} dx  
\frac{x^{\alpha}(x+\mu/T)^{3/2}e^x}{(e^x +1)^2},
\label{Lb}
\end{equation} 
where $c$ is the number of the carrier pockets and we take $c=1$,
$m^{*}=0.02\,m_e$ is the effective mass, and temperature is measured
in units of $k_B$. We estimate $\tau = 10^{-14}$ s. These $L_\alpha$
are then substituted into Eq.~\eqref{ZT2} to give $ZT$ which is shown
in Fig.~\ref{fig:ZT2} for a range of densities $n$ and chemical
potential $\mu$ at room temperature.  The maximum, for the estimated
parameters, is higher than 6, which makes these systems quite unique
even if something of that order can be reached.

We note that, if the impurities are non-magnetic, the transition to a
bulk Anderson insulator should not destroy the topologically protected
1D states, because the time reversal invariance is not broken. Also,
note that by increasing number of dislocations or non-magnetic
impurities one can come closer to the 1D wire limit for $ZT$ and reach
even greater values. These densities are however more unlikely to be
achieved and can lead also to tunneling between the channels which
could lead to an opening of a gap and localization of the protected
states. This is the reason for our choice of an upper limit of the
1D-channel density of $n\sim 10^{17} {\rm m}^{-2}$ corresponding to a
typical spacing of $\sim 3$ nm.

\textit{Summary.}  We have studied the thermoelectric properties of 3D
topological insulators with non-zero TRIM which contain many line
dislocations possessing topologically-protected perfectly conducting
1D states. We have shown that in principle this system can have very
high figure of merit, $ZT\sim 10$, and predict that increasing the
number of dislocations in a TI film will exhibit an increase in $ZT$.

This work was supported by NSF under Grant No. DMR-0547875 and Grant
No. 0757992, by the Research Corporation Cottrell Scholar Award, and
by the Welch Foundation (A-1678). J.~S. and S.~M.  thank the
KITPC-Beijing for their kind hospitality during which part of this
work was done (activity name ``Progress in Spintronics and Graphene
Research'').

\bibliography{thermoelectrictopological}
\end{document}